\documentclass[conference]{IEEEtran}
\IEEEoverridecommandlockouts

\usepackage{cite}
\usepackage{amsmath,amssymb,amsfonts}
\usepackage{algorithmic, algorithm}
\usepackage{graphicx}
\usepackage{textcomp}
\usepackage{xcolor}
\usepackage{amsthm}
\usepackage{multirow}
\usepackage{enumitem}
\usepackage{color}
\usepackage{needspace}
\usepackage{array}
\usepackage{stfloats}
\usepackage{url}
\usepackage{balance}
\usepackage{caption}
\usepackage{booktabs,microtype}
\usepackage{tikz}
\usepackage{subcaption}
\usetikzlibrary{positioning}
\renewcommand{\baselinestretch}{0.92}
\definecolor{seamblue}{RGB}{34,91,141}
\newcommand{\red}[1]{#1}
\newcommand{\pos}[1]{\left[#1\right]_{+}}
\newcommand{\blue}[1]{#1}
\newcommand{\yl}[1]{#1}
\newcommand{\green}[1]{#1}

\def\BibTeX{{\rm B\kern-.05em{\sc i\kern-.025em b}\kern-.08em
    T\kern-.1667em\lower.7ex\hbox{E}\kern-.125emX}}
\begin{document}

\title{Reverse Seam-Carving: Structured Whitespace Preconditioning for Fixed-Die P\&R}
\author{
Andrew B. Kahng,
Yiting Liu,
and Davit Markarian\\
\{abk, yil375, dmarkarian\}@ucsd.edu
\medskip
\\
\small  
UC San Diego, La Jolla, CA, USA \\
\vspace{-2.5mm}%
}
\maketitle


\begin{abstract}

At high placement utilization, limited whitespace
increases cell displacement during legalization and
clock tree synthesis (CTS), degrading routed wirelength.
Existing P\&R methodologies use uniform density
targets or reactive whitespace redistribution, without
optimizing whitespace structure before placement.
We propose Reverse Seam-Carving (RSC), a \yl{pre-placement}
methodology that redistributes whitespace into core-spanning
low-density seams between denser regions, while preserving
die area and utilization.
We further develop \emph{SeamGain}, a \blue{semi-analytical model} that selects
the RSC configuration \red{from a single reference run of the uniform flow};
one \emph{SeamGain}-selected run obtains 86\% of the
best-of-sweep wirelength reduction.
In the OpenROAD flow with ASAP7, RSC achieves
(maximum, average) post-route wirelength reductions of
(27.6\%, 15.4\%).

\end{abstract}


\section{Introduction}
\label{sec:introduction}

Placement quality significantly impacts routed wirelength
and routability in VLSI physical design.
In a fixed-die implementation at high placement utilization,
limited whitespace leads to increased cell displacement
during legalization, routing congestion, and degraded
power, performance and area (PPA).
By contrast, variable-die place-and-route \yl{(P\&R)} can ``flex'',
i.e., create extra room wherever a hotspot is
encountered~\cite{VujkovicWSS04}.
Furthermore, PROBE~\cite{KahngLL18} observes that, at the
same utilization, fatal hotspots are more likely in larger
layout regions.
These observations motivate a key question: can we obtain
the benefit of \blue{variable-die} flex within a fixed die, by deliberately
organizing the existing whitespace between smaller and
denser placement regions, without increasing die area?

State-of-the-art analytical placers optimize wirelength
subject to cell density constraints~\cite{replace}, and
existing methodologies employ various
whitespace allocation strategies to improve routability.
Free-space management distributes whitespace during
placement~\cite{AlpertNV03}.
Congestion-driven approaches redistribute whitespace
according to estimated routing demand or
placement and routing feedback~\cite{YangCS02}\cite{LiXKCM07}.
Other approaches use cell inflation or learned density
targets to guide placement~\cite{goalplace}\cite{replace}.
However, these approaches often require placement and
routing iterations (i.e., loops) to identify
congestion hotspots and reactively
insert density screens to alleviate local congestion.
Moreover, it is not known in advance whether a given
whitespace structure helps a given design, so that
candidate structures must be evaluated by complete
P\&R runs.

\begin{figure}[t]
    \centering
    \includegraphics[width=\columnwidth]{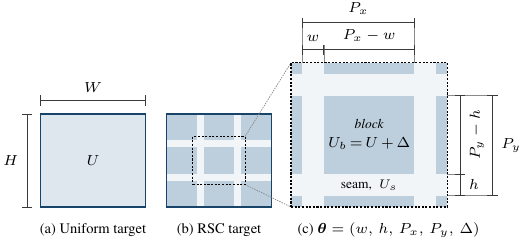}
\caption{(a) Uniform density target $U$ over a $W \times H$ core. (b) RSC
redistributes whitespace into core-spanning seams at fixed core area and
overall utilization. (c) An RSC configuration $\boldsymbol{\theta}$:
vertical seams of width $w$ at pitch $P_x$ and horizontal seams of height
$h$ at pitch $P_y$ separate blocks with \yl{density} target $U_b = U + \Delta$; seams
have a lower \yl{density} target $U_s$.}
    \label{fig:rsc}
    \vspace{-1em}
\end{figure}

We propose \emph{Reverse Seam-Carving} (RSC),
a structured \yl{pre-placement} whitespace allocation
methodology that significantly reduces routed wirelength
at high utilization.
Inspired by seam-carving in image processing~\cite{AvidanS07},
RSC redistributes whitespace into core-spanning horizontal
and vertical low-density seams that separate denser
placement regions, while preserving the overall utilization
and die area (see Figure~\ref{fig:rsc}).
The seams bring the flex of variable-die implementation
into a fixed die: they hold whitespace within a short
distance of every cell, so that legalization and clock
tree synthesis (CTS) absorb cell displacement and clock
buffers locally, instead of degrading the wirelength
achieved by global placement.
RSC is also simple to deploy,\footnote{\green{RSC adds only
32 lines of Tcl to the baseline (see \texttt{seams.tcl} in~\cite{rsc_repo}).}} 
\blue{as it} 
preserves the flat netlist and requires no changes to
the placer or the router.
We further develop \emph{SeamGain}, a 
\blue{semi-analytical} model
that automatically
selects the seam configuration for a given design.
Our main contributions are as follows.
\begin{itemize}[noitemsep, topsep=0pt, leftmargin=*]

\item We propose \emph{RSC}, a proactive
\yl{pre-placement} methodology that organizes whitespace
into connected horizontal and vertical low-density
seams under a fixed die-area and utilization budget.
RSC reduces routed wirelength by providing legalization
and CTS with nearby whitespace, and requires only
density screens in an existing P\&R flow.

\item We develop \emph{SeamGain}, \blue{a physically motivated semi-analytical model that} automatically
selects the seam configuration from \green{a single
reference run of the uniform flow (i.e., the flow without seams), }
without any RSC run or routing.
\emph{SeamGain} also decides whether RSC should be
applied to a design, and thus replaces a costly
parameter sweep.

\item We evaluate RSC using six ASAP7 designs in the
open-source OpenROAD flow~\cite{openroad}.
RSC achieves (maximum, average) post-route wirelength
reductions of (27.6\%, 15.4\%), reduces congestion and
detailed routing CPU time by 7.8\% and 25.6\% on average, and
\blue{completes cases in which the uniform flow fails}.
A single \emph{SeamGain}-selected RSC run achieves 86\%
of the best-of-144 sweep improvement
\blue{(94\% when the best
swept configurations are rerun with other placement seeds)}.

\end{itemize}

The remaining sections are organized as follows.
Section~\ref{sec:problem} introduces the terminology
and the problem formulation.
Section~\ref{sec:approach} discusses our approach.
Section~\ref{sec:experiments} shows experimental results,
and Section~\ref{sec:conclusion} concludes the paper.
\yl{All code, scripts and flow settings are publicly available~\cite{rsc_repo}.}

\section{Problem Formulation}
\label{sec:problem}

This section presents the RSC configuration, the
whitespace budget and the optimization objective.
All terms used in this paper are summarized in
Table~\ref{tab:notations}.

\begin{table}[!t]
\caption{Terminology and notation.}
\label{tab:notations}
\centering
\begingroup
\scriptsize
\setlength{\tabcolsep}{3pt}
\renewcommand{\arraystretch}{1.03}
\begin{tabular}{|p{1.4cm}|p{7.1cm}|}
\hline
\textbf{Notation} & \textbf{Description} \\ \hline
Uniform flow & The P\&R flow with a single density target $U$ and no seams; it is the baseline in our experiments. \\ \hline
Reference run & A single run of the uniform flow through placement and CTS, without routing; it provides the inputs of \emph{SeamGain}. \\ \hline

\multicolumn{2}{|l|}{\textbf{RSC configuration and whitespace}} \\ \hline
$W$, $H$ & Core width and height. \\ \hline
$U$, $U_b$, $U_s$ & Overall utilization; \yl{block and seam density targets}. \\ \hline
$\Delta$ & \green{Increase of the block density target over $U$}, $U_b-U$. \\ \hline
$w$, $h$ & Vertical-seam width (sites) and horizontal-seam height (rows). \\ \hline
$P_x$, $P_y$ & Vertical-seam pitch along $x$ (sites) and horizontal-seam pitch along $y$ (rows). \\ \hline
$\boldsymbol{\theta}$, $\Theta$ & RSC configuration $(w,h,P_x,P_y,\Delta)$ and set of candidate configurations. \\ \hline
$f$ & Fraction of placement area covered by seams. \\ \hline
$W_s$ & Seam whitespace as a fraction of placement area. \\ \hline
$\delta$ & Cell packing density after excluding seam whitespace, $U/(1-W_s)$. \\ \hline
\multicolumn{2}{|l|}{\textbf{Wirelength decomposition}} \\ \hline
$L_{\rm gp}$ & HPWL after global placement. \\ \hline
$L_{\rm rWL}(\boldsymbol{\theta})$ & Routed wirelength with configuration $\boldsymbol{\theta}$. \\ \hline
$\lambda_{\rm dp}$, $\lambda_{\rm cts}$ & \blue{HPWL after/before legalization, and after/before the legalization that follows CTS.} \\ \hline
$\rho$ & Routed wirelength divided by HPWL after CTS legalization. \\ \hline
$\Delta L_{\rm gp}$ & Relative change in global-placement HPWL versus the uniform \yl{flow}. \\ \hline
$c_b$, $c_s$, $\bar{c}$ & Mean number of nets crossing a cut line in blocks, seams, and the core. \\ \hline
\multicolumn{2}{|l|}{\textbf{SeamGain prediction model}} \\ \hline
$e_{\rm dp}$, $e_{\rm cts}$ & \green{Excess inflations of the reference run}, $\lambda_{\rm dp}-1$ and $\lambda_{\rm cts}-1$. \\ \hline
$a_{\rm cts}$ & Clock-buffer area divided by placement area in the \green{reference} run. \\ \hline
$R_{\rm dp}$, $R_{\rm cts}$ & Recovered fractions of the DP and CTS excess inflations. \\ \hline
$p$, $P_0$ & Relative pitch $p=P_x/P_0-1$; reference pitch $P_0=300$ sites. \\ \hline
$r_d$, $r_c$ & DP and CTS baseline recovery coefficients. \\ \hline
$g_d$, $g_c$ & Sensitivity of DP and CTS recovery to seam pitch. \\ \hline
$k_d$, $k_c$ & DP and CTS packing-penalty coefficients. \\ \hline
$\delta_0$ & Packing density above which recovery penalties apply. \\ \hline
$\kappa$ & Correction for changes in $\rho L_{\rm gp}$ relative to the uniform \yl{flow}. \\ \hline
$\widehat{G}(\boldsymbol{\theta})$ & Predicted wirelength reduction versus the uniform \yl{flow}. \\ \hline
\multicolumn{2}{|l|}{\textbf{Configuration selection}} \\ \hline
$\varnothing$ & Uniform \yl{flow} (no RSC). \\ \hline
$\boldsymbol{\theta}^{*}$ & Wirelength-minimizing choice among RSC candidates and \yl{the uniform flow}. \\ \hline
$\delta_{\max}$ & Maximum allowed packing density for candidate selection. \\ \hline
$\tau$, $\epsilon$ & Minimum predicted reduction to apply RSC; near-best prediction tolerance. \\ \hline
$\Theta_f$, $\Theta_t$ & Feasible configurations; near-best configurations for tie-breaking. \\ \hline
\end{tabular}
\endgroup
\vspace{-1.5em}
\end{table}

\subsection{RSC Configuration}

Given a netlist, floorplan, and target utilization,
RSC redistributes placement whitespace into
core-spanning horizontal and vertical low-density
seams. The seams divide the placement canvas into
denser and smaller placement regions, referred to as
\emph{blocks}, without partitioning
the netlist or changing the die area.
We characterize an RSC configuration by
\begin{equation}
\boldsymbol{\theta} =
(w,h,P_x,P_y,\Delta),
\label{eq:rsc_config}
\end{equation}
where $w$ and $h$ denote the width of vertical
seams and height of horizontal seams, respectively;
$P_x$ and $P_y$ denote their corresponding pitches;
and $\Delta$ denotes the increase \yl{of the block
density target}
relative to the overall utilization.

\subsection{Whitespace Redistribution}

Let $f$ denote the fraction of the placement area
covered by the seams.
For an ideal regular grid,
\begin{equation}
f \approx 1-
\left(1-\frac{w}{P_x}\right)
\left(1-\frac{h}{P_y}\right).
\label{eq:seam_fraction}
\end{equation}
In practice, $f$ is computed from the available
placement sites, with the seams clipped at the
core boundary.
Let $U$, $U_s$, and $U_b$ denote the overall
\yl{utilization, and the seam and block density
targets,}
respectively. To preserve the overall
utilization, the density targets satisfy
$U=(1-f)U_b+fU_s$.
With $\Delta=U_b-U$, we obtain
\begin{equation}
U_s=U-\frac{1-f}{f}\Delta.
\label{eq:seam_util}
\end{equation}
Equation~\eqref{eq:seam_util} shows that a
small increase of the \green{block density target
substantially reduces the seam density target} when the
seam-area fraction is small.
The whitespace held by the seams, as a fraction of
the placement area, is
\begin{equation}
W_s = f(1-U_s) = f(1-U)+(1-f)\Delta.
\label{eq:seam_ws}
\end{equation}
Since the seams reserve $W_s$, the cells are packed
into the remaining area at a density of
\begin{equation}
\delta = \frac{U}{1-W_s}.
\label{eq:packing}
\end{equation}

\subsection{Optimization Objective}

The RSC configuration problem is defined as follows.
We are given:
\begin{itemize}[noitemsep, topsep=0pt, leftmargin=*]
\item a design with a fixed netlist, technology
and timing constraints;
\item a fixed-die floorplan with overall utilization $U$
and fixed I/O locations; and
\item a set $\Theta$ of candidate RSC configurations.
\end{itemize}
Let $L_{\rm rWL}(\boldsymbol{\theta})$
denote the routed wirelength obtained using
RSC configuration $\boldsymbol{\theta}$,
and let $L_{\rm rWL}(\varnothing)$ denote that of the
uniform \yl{flow}.
The objective is to identify
\begin{equation}
\boxed{
\boldsymbol{\theta}^{*}
=
\arg\min_{\boldsymbol{\theta}\in\Theta\cup\{\varnothing\}}
L_{\rm rWL}(\boldsymbol{\theta}),
}
\label{eq:rsc_objective}
\end{equation}
subject to legal placement and routing completion.
Since evaluating each candidate through a complete
P\&R flow is expensive, our goal is to identify an
effective configuration, or to keep the \yl{uniform flow},
without exhaustive P\&R evaluation.

\section{Our Approach}
\label{sec:approach}

\begin{figure}[t]
    \centering
    \includegraphics[width=\columnwidth]{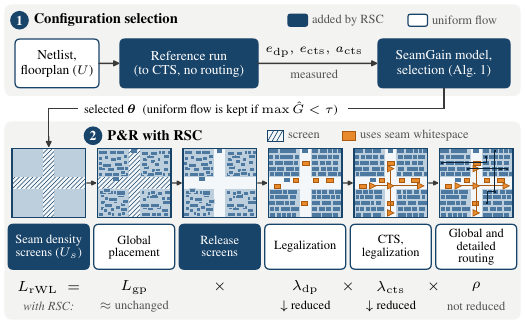}
\caption{The RSC flow. Dark boxes are the steps added to the uniform
flow. (1) The reference run provides the inputs of the SeamGain model,
which selects the configuration $\boldsymbol{\theta}$ or keeps the uniform
flow. (2) Seams are realized as density screens during global placement and
released before legalization, so that displaced cells and clock buffers
(orange) use the seam whitespace.}
    \label{fig:flow}
    \vspace{-1em}
\end{figure}

Figure~\ref{fig:flow} shows our approach.
\green{We refer to the P\&R flow with a single density
target $U$ and no seams as the \emph{uniform flow}.
Given a design and a floorplan, we first run the uniform
flow once, through placement and
CTS, without routing.
This single run is the \emph{reference run}; it}
provides the design-dependent
inputs of \emph{SeamGain}
(Section~\ref{subsec:model}), which selects
the RSC configuration (Section~\ref{subsec:selection}).
The selected configuration is then realized as density
screens before global placement.

\begin{figure}[t]
    \centering
    \subfloat[Uniform \yl{flow}]{\includegraphics[width=0.38\columnwidth]{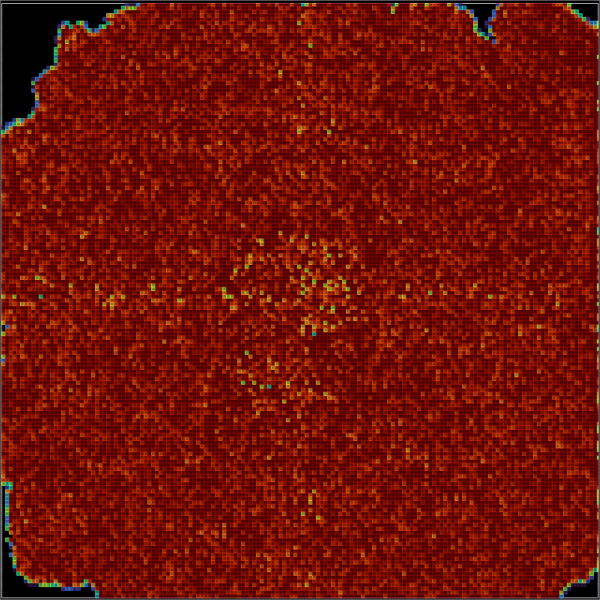}\label{fig:density_uniform}}
    \hfil
    \subfloat[RSC]{\includegraphics[width=0.38\columnwidth]{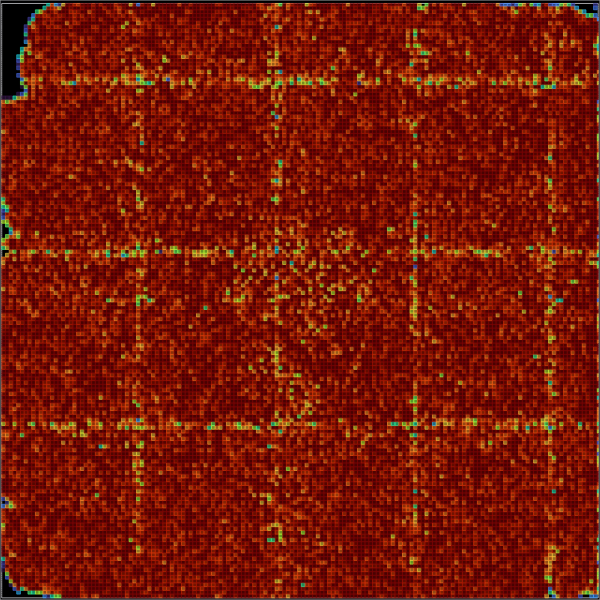}\label{fig:density_rsc}}
\caption{\blue{Placement density of \textit{jpeg} at $U=91\%$ with
\yl{(a) the uniform flow and (b) RSC. }}}
    \label{fig:density}
    \vspace{-1em}
\end{figure}

\subsection{RSC in the P\&R Flow}
\label{subsec:flow}

Each seam is realized as a partial placement blockage
(i.e., a \emph{density screen}) with maximum density $U_s$
from Equation~\eqref{eq:seam_util}.
The first seam in each direction is placed one pitch
from the core origin.
Configurations with $U_s$ below 5\% require hard
placement blockages and are not considered.
Global placement is then performed under the
density screens.
\green{Figure \ref{fig:density} shows the placement density of \textit{jpeg}
at $U=91\%$ with the uniform flow and with RSC, where
seams have visibly lower density (i.e., site occupancy).}
The screens are released before legalization, so that
the whitespace held by the seams is available to
legalization and to the buffers inserted by CTS.
Subsequent steps correspond to the uniform flow:
legalization, CTS, \green{the legalization that follows CTS},
global routing and detailed routing.
RSC changes neither the die area nor the routing
resources.

\subsection{How RSC Affects Routed Wirelength}
\label{subsec:mechanism}

We decompose routed wirelength by flow stage as
\begin{equation}
L_{\rm rWL} \approx \rho\cdot\lambda_{\rm cts}\cdot\lambda_{\rm dp}\cdot L_{\rm gp},
\label{eq:stages}
\end{equation}
where $L_{\rm gp}$ is the half-perimeter wirelength (HPWL)
at the end of global placement;
\blue{$\lambda_{\rm dp}$ is the \emph{inflation factor} of
legalization, i.e., the HPWL after legalization divided by
the HPWL before it;
$\lambda_{\rm cts}$ is the same ratio for the legalization
that follows CTS; and
$\rho$ is the \emph{routing factor}, i.e., the routed
wirelength divided by the HPWL after the CTS legalization.}
RSC affects the four factors differently.

\noindent\textbf{Global placement ($L_{\rm gp}$). }
Redistributing whitespace compresses the blocks
by a factor of $U/U_b$ and stretches the seams
by a factor of $U/U_s$.
Since the wirelength along one axis is the integral
of the number of nets crossing each cut line,
the first-order relative change of wirelength is
\begin{equation}
\Delta L_{\rm gp} \approx
-(1-f)\,\frac{\Delta}{U}\cdot\frac{c_b-c_s}{\bar{c}},
\label{eq:cut}
\end{equation}
where $c_b$, $c_s$ and $\bar{c}$ denote the average
number of nets crossing a cut line inside the blocks,
inside the seams and over the core, respectively.
For a regular seam grid, the seams are ordinary cut lines,
i.e., $c_s\approx c_b$.
Hence, RSC is not expected to change the global
placement wirelength.

\noindent\textbf{Legalization and CTS ($\lambda_{\rm dp}$, $\lambda_{\rm cts}$). }
At high utilization, legalization must displace cells from locally 
overfilled regions to distant free placement sites, and CTS inserts 
clock buffers into regions that are already densely occupied. Both 
effects increase the HPWL achieved by global placement. RSC reserves a 
whitespace of $W_s$ in the seams, such that free placement sites are
available within one seam pitch of every cell. This reduces the cell 
displacement required by legalization and by clock buffer insertion, 
thereby reducing $\lambda_{\rm dp}$ and $\lambda_{\rm cts}$.
On the other hand, RSC also increases the packing density of the blocks 
from $U$ to $\delta$, which makes legalization within the blocks more 
difficult.

\noindent\textbf{Routing ($\rho$). }
RSC does not add routing resources. Hence,
RSC is not expected to reduce routing detours.

Therefore, RSC can only recover the wirelength that the
uniform flow loses in legalization and CTS, i.e., its
reduction is bounded by
$1-1/(\lambda_{\rm dp}\lambda_{\rm cts})$ of the
uniform flow. The validation of this analysis
is shown in Section~\ref{subsec:source}.

\subsection{\blue{SeamGain Model}}
\label{subsec:model}

Based on the above analysis, we develop
\emph{SeamGain}, a model that predicts the
routed wirelength reduction of an RSC
configuration $\boldsymbol{\theta}$ relative to the
\yl{uniform flow}.
The key idea is to estimate how much of the
wirelength inflation introduced during detailed
placement (DP) and CTS
can be avoided by RSC.

\noindent\textbf{Recovered inflation. }
Let $e_{\rm dp}=\lambda_{\rm dp}-1$ and
$e_{\rm cts}=\lambda_{\rm cts}-1$ denote the
\emph{excess inflations} of the two stages in
the reference run.
We define the \emph{recoveries} $R_{\rm dp}$ and
$R_{\rm cts}$ as the fractions of these excess
inflations that RSC removes.
Thus, the inflation factors with RSC are
estimated as $1+(1-R_{\rm dp})e_{\rm dp}$ and
$1+(1-R_{\rm cts})e_{\rm cts}$: a recovery of
one removes the inflation of a stage, and a
recovery of zero leaves it unchanged.
We model the recoveries as functions of the
seam whitespace $W_s$, the seam pitch $P_x$ and
the packing density $\delta$ of the configuration:
\begin{align}
R_{\rm dp}(\boldsymbol{\theta}) &=
  r_d(1-g_d p)
  - k_d\pos{\delta-\delta_0},
  \label{eq:rdp} \\
R_{\rm cts}(\boldsymbol{\theta}) &=
  r_c(1-g_c p)
  \min\!\left\{1,\frac{W_s}{a_{\rm cts}}\right\}
  - k_c\pos{\delta-\delta_0},
  \label{eq:rcts}
\end{align}
where $p=P_x/P_0-1$ is the relative seam
pitch ($P_0=300$ sites),
$\pos{x}=\max\{0,x\}$, and
$a_{\rm cts}$ is the clock-buffer area of the reference run, normalized by the
placement area.
The coefficients $r_d$ and $r_c$ are the baseline
recoveries, i.e., the recoveries at the reference
pitch ($p=0$) when the seam whitespace is
sufficient and the blocks are not over-packed.
The other terms capture three effects.
(i) \emph{Seam headroom}: CTS inserts clock
buffers with a total area of $a_{\rm cts}$.
The factor $\min\{1,W_s/a_{\rm cts}\}$ scales the
CTS recovery with the seam whitespace $W_s$, and
saturates once the seams can hold all clock
buffers ($W_s\geq a_{\rm cts}$).
(ii) \emph{Seam pitch}: a smaller pitch places
cells closer to the seams. The factors
$(1-g_d\,p)$ and $(1-g_c\,p)$ increase both
recoveries as $P_x$ decreases.
(iii) \emph{Block packing}: when the packing
density $\delta$ (Equation~\eqref{eq:packing})
exceeds a threshold $\delta_0$, legalization
within the blocks becomes more difficult.
The penalties $k_d\pos{\delta-\delta_0}$ and
$k_c\pos{\delta-\delta_0}$ reduce both recoveries,
which may become negative, i.e., RSC then
increases the inflation.

\noindent\textbf{Predicted reduction. }
By Equation~\eqref{eq:stages}, routed wirelength
is the product of $\rho L_{\rm gp}$ and the two
inflation factors.
We therefore predict the wirelength reduction of
$\boldsymbol{\theta}$ as
\begin{equation}
\widehat{G}(\boldsymbol{\theta}) =
1-\kappa
\frac{
[1+(1-R_{\rm dp})e_{\rm dp}]
[1+(1-R_{\rm cts})e_{\rm cts}]
}{
(1+e_{\rm dp})(1+e_{\rm cts})
},
\label{eq:model}
\end{equation}
where the fraction is the estimated ratio of
$\lambda_{\rm dp}\lambda_{\rm cts}$ between RSC and the
\yl{uniform flow}, and $\kappa$ is the ratio of
$\rho L_{\rm gp}$, i.e., it accounts for the changes
in global-placement wirelength and in the routing
factor that the recoveries do not capture.

\noindent\textbf{Inputs and calibration. }
\emph{SeamGain} has three design-dependent
inputs, $(e_{\rm dp},e_{\rm cts},a_{\rm cts})$,
which are measured in the
reference run, and eight parameters,
$(r_d,g_d,k_d,r_c,g_c,k_c,\delta_0,\kappa)$,
which are shared by all designs.
We calibrate the parameters using RSC runs of
training designs: we fit
Equations~\eqref{eq:rdp} and~\eqref{eq:rcts}
to the measured DP and CTS recoveries, and set
$\kappa$ to the median ratio of
$\rho L_{\rm gp}$ between RSC and the uniform flow.
For an unseen design, \emph{SeamGain} 
predicts the wirelength reduction of any
candidate configuration from its parameters
and the reference run of the design,
without requiring an RSC run.

\subsection{Design-Aware Configuration Selection}
\label{subsec:selection}

Algorithm~\ref{alg:select} summarizes the selection
of the RSC configuration.
We first discard the configurations whose packing density
exceeds a packing limit $\delta_{\max}$, beyond which
legalization may fail (Lines 2--3).
We then predict the wirelength reduction of the remaining
configurations by Equation~\eqref{eq:model} (Line 4).
RSC is applied only if the best predicted reduction is at
least a threshold $\tau$; otherwise, the \yl{uniform flow}
is kept (Lines 5--7).
\blue{The complete procedure thus requires one reference
run through CTS and at most one RSC run.}
Since several configurations may have similar predicted reductions, we select, among those within $\epsilon$ of the best prediction, the configuration whose seam whitespace is closest to the midpoint of the feasible interval $[a_{\rm cts},\,1-U/\delta_{\max}]$. This maximizes the minimum margin to the headroom and packing limits (Lines 8--9).

\begin{algorithm}[t]
\caption{Design-aware RSC configuration selection.}
\label{alg:select}
\footnotesize
\begin{algorithmic}[1]
\REQUIRE Design at utilization $U$, candidate configurations $\Theta$, \emph{SeamGain} \yl{parameters}, $\delta_{\max}$, $\tau$, $\epsilon$
\ENSURE RSC configuration $\boldsymbol{\theta}$, or \green{perform uniform flow}
\STATE Run the uniform flow through CTS; measure $e_{\rm dp}$, $e_{\rm cts}$ and $a_{\rm cts}$
\STATE Compute $U_s$, $W_s$ and $\delta$ of each $\boldsymbol{\theta}\in\Theta$ by Equations~\eqref{eq:seam_util}--\eqref{eq:packing}
\STATE $\Theta_f\leftarrow\{\boldsymbol{\theta}\in\Theta: U_s\geq5\%,\ \delta\leq\delta_{\max}\}$
\STATE Compute $\widehat{G}(\boldsymbol{\theta})$ for each $\boldsymbol{\theta}\in\Theta_f$ by Equation~\eqref{eq:model}
\IF{$\Theta_f=\emptyset$ \OR $\max_{\Theta_f}\widehat{G}<\tau$}
\RETURN \green{\texttt{// perform uniform flow}}
\ENDIF
\STATE $\Theta_t\leftarrow\{\boldsymbol{\theta}\in\Theta_f:\widehat{G}(\boldsymbol{\theta})\geq\max_{\Theta_f}\widehat{G}-\epsilon\}$
\RETURN $\boldsymbol{\theta}\in\Theta_t$ with $W_s$ closest to $\frac{1}{2}\bigl(a_{\rm cts}+1-U/\delta_{\max}\bigr)$
\end{algorithmic}
\end{algorithm}

\section{Experimental Validation}
\label{sec:experiments}

RSC is implemented on the OpenROAD
infrastructure~\cite{openroad} (commit hash: ab8ddb7).
We evaluate RSC using six ASAP7~\cite{asap7} designs.
Table~\ref{tab:testcases} summarizes the characteristics
of these benchmarks.
All netlists are synthesized using Cadence Genus 25.1,
including buffering and sizing for timing closure.
Each netlist is repaired once before floorplanning and
is fixed afterward; the only netlist changes during P\&R
are the buffers and nets inserted by CTS.
Floorplans have an aspect ratio of 1, and I/O placement
is fixed after floorplanning.
We use routability-driven global placement, routing layers
M1--M6, and a 25\% reduction of the global routing capacity.
For a given design and utilization \green{(i.e., a case)}, the uniform
flow, \green{which is our baseline}, and all RSC configurations use the same
netlist, floorplan and I/O locations.

We sweep the utilization $U\in\{90\%, 91\%, 92\%, 93\%\}$.
For \textit{tpu}, \textit{ethmac} and \textit{jpeg}, we sweep
$w\in\{10,15\}$ sites, $h\in\{2,3\}$ rows,
$\Delta\in\{1\%,\dots,6\%\}$ and six pitches $(P_x,P_y)$
from $(240,60)$ to $(360,90)$ $(P_y=P_x/4)$, i.e., 144 configurations
per design and utilization.
For \textit{cva6}, \textit{swerv} and \textit{mempool}, we sweep
$\Delta\in\{2\%,4\%,6\%\}$ and three pitches from
$(280,70)$ to $(320,80)$, i.e., 36 configurations.
\yl{The uniform flow} is run with five global placement seeds.
RSC configurations are \yl{evaluated} by global routing with
one \yl{placement} seed; the best configurations per design and
utilization are rerun with five seeds and detail-routed.
Wirelength reductions are reported against the median of
the baseline seeds.

\begin{table}[!htb]
\caption{Characteristics of our benchmarks. $a_{\rm cts}$ is measured
in the reference runs (median over $U=90\%$--$92\%$).}
\label{tab:testcases}
\centering
\scriptsize
\setlength{\tabcolsep}{4.5pt}
\begin{tabular}{|l|r|r|r|r|r|}
\hline
\textbf{Design} & \textbf{\#Insts} & \textbf{\#Nets} & \textbf{\#IOs} & \textbf{Area ($\mu m^2$)} & $a_{\rm cts}$ \\ \hline
\textit{tpu}            & 35,636 & 35,768 & 389   & 3,770.59  & 0.97\% \\ \hline
\textit{ethmac}         & 36,002 & 36,102 & 216   & 5,176.85  & 3.94\% \\ \hline
\textit{jpeg}           & 46,790 & 46,811 & 47    & 4,900.96  & 0.75\% \\ \hline
\textit{cva6}           & 67,627 & 71,653 & 5,257 & 7,521.49  & 1.02\% \\ \hline
\textit{swerv\_wrapper} & 85,567 & 86,380 & 1,416 & 10,804.74 & 0.58\% \\ \hline
\textit{mempool\_tile}  & 95,991 & 96,819 & 1,316 & 11,742.88 & 0.69\% \\ \hline
\end{tabular}
\end{table}

In this section, we first present the post-route results
(Section~\ref{subsec:postroute}) and the source of the
wirelength reduction (Section~\ref{subsec:source}).
We then evaluate the accuracy of \emph{SeamGain}
(Section~\ref{subsec:accuracy}) and the configuration
selection (Section~\ref{subsec:selection_eval}).

\subsection{Post-Route Results}
\label{subsec:postroute}

\begin{table}[!htb]
\caption{Post-route metrics of the best RSC configuration, normalized to the \yl{uniform flow}
(medians over \red{DRC-}clean detailed routing runs). rWL: routed wirelength of signal nets.
Cong.: maximum layer demand over capacity in global routing. CPU: detailed routing CPU time.
Seeds: number of \red{DRC-}clean (baseline, RSC) runs.}
\label{tab:main_results}
\centering
\scriptsize
\setlength{\tabcolsep}{3.4pt}
\begin{tabular}{|l|c|c|c|c|c|c|c|c|}
\hline
\textbf{Design} & $U$ & $P_x$ & $\Delta$ & \textbf{HPWL} & \textbf{rWL} & \textbf{Cong.} & \textbf{CPU} & \textbf{Seeds} \\ \hline
\multirow{4}{*}{\textit{tpu}} & 90\% & 300 & 2\% & \textbf{0.835} & \textbf{0.843} & \textbf{0.896} & 0.886 & (5, 5) \\ \cline{2-9}
 & 91\% & 240 & 2\% & \textbf{0.813} & \textbf{0.827} & \textbf{0.883} & 0.819 & (5, 5) \\ \cline{2-9}
 & 92\% & 240 & 2\% & \textbf{0.789} & \textbf{0.794} & \textbf{0.872} & 0.753 & (5, 1) \\ \cline{2-9}
 & 93\% & 240 & 1\% & \textbf{0.829} & \textbf{0.816} & 1.024 & 0.674 & (5, 1) \\ \hline
\multirow{2}{*}{\textit{ethmac}} & 90\% & 240 & 3\% & \textbf{0.809} & \textbf{0.820} & \textbf{0.919} & 0.639 & (2, 5) \\ \cline{2-9}
 & 91\% & 240 & 2\% & \textbf{0.769} & \textbf{0.724} & \textbf{0.855} & 0.458 & (1, 1) \\ \hline
\multirow{4}{*}{\textit{jpeg}} & 90\% & 240 & 2\% & \textbf{0.908} & \textbf{0.912} & \textbf{0.940} & 0.562 & (5, 5) \\ \cline{2-9}
 & 91\% & 320 & 2\% & \textbf{0.900} & \textbf{0.904} & \textbf{0.929} & 0.899 & (5, 5) \\ \cline{2-9}
 & 92\% & 320 & 2\% & \textbf{0.905} & \textbf{0.911} & \textbf{0.941} & 0.863 & (5, 5) \\ \cline{2-9}
 & 93\% & 260 & 1\% & \textbf{0.905} & \textbf{0.907} & \textbf{0.965} & 0.885 & (5, 5) \\ \hline
\end{tabular}
\end{table}

We now present the post-route results of RSC.
For each design and utilization, we select the
detail-routed configuration with the smallest median
routed wirelength (rWL), and compare it against the
\yl{uniform flow} in Table~\ref{tab:main_results}.
We report the ten cases of \textit{tpu}, \textit{ethmac} and \textit{jpeg} for which
both the baseline and RSC have \red{DRC-}clean detailed routing
results.\footnote{The other three designs have no \red{DRC-}clean
pair of detailed routing results, except \textit{swerv} at $U=90\%$
(2.0\% reduction, within its seed variation); we analyze
them using global routing.}
RSC achieves (maximum, average) rWL reductions of
(27.6\%, 15.4\%); the average reductions are 18.0\%,
22.8\% and 9.2\% for \textit{tpu}, \textit{ethmac} and \textit{jpeg}, respectively.
Congestion and detailed routing CPU time are
reduced by 7.8\% and 25.6\% on average.\footnote{\red{RSC also 
leads to clear improvements in post-route timing (WNS and TNS), along with
total power reductions, for all ten cases; see \texttt{data/} in~\cite{rsc_repo}.}}
The best configurations use $\Delta$ between 1\% and 3\%,
and seven of the ten use the two smallest pitches.

These reductions are larger than the variation across
placement seeds.
When RSC is restricted to configurations with five \red{DRC-}clean
\yl{runs}, the average rWL reduction is 14.4\%.
For the 34 configurations of \textit{tpu} and \textit{jpeg} with at least
four \red{DRC-}clean \yl{runs} at $U=90\%$--$92\%$, the rWL reductions
are 14.4\%--18.7\% and 7.3\%--9.6\%, respectively, and
the worst RSC \yl{run} has at least 3.5\% smaller rWL than the
best baseline run.
Moreover, at $U=90\%$--$92\%$, 59\% to 100\% of the \yl{swept}
configurations of \textit{tpu}, \textit{ethmac} and \textit{jpeg} that complete global
routing have smaller wirelength than the best
of the five baseline seeds.

RSC also improves flow completion.
For \textit{ethmac} at $U=92\%$, the \yl{uniform flow} does not
produce a \red{DRC-}clean detailed routing result within the
evaluated runs, whereas 13 RSC runs complete detailed
routing without violations.
For \textit{ethmac} at $U=93\%$, the uniform flow fails the
legalization after CTS for all five seeds, whereas 36 of
the 48 RSC configurations with $\Delta=4\%$ or $5\%$
complete global routing. 

\begin{figure*}[t]
  \centering
  \includegraphics[width=0.85\textwidth]{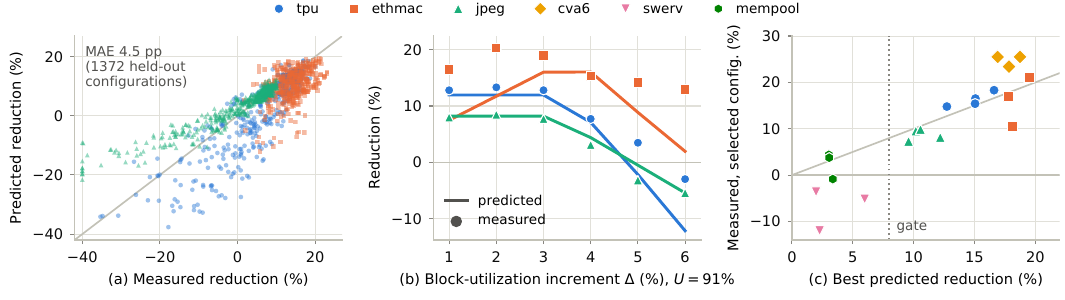}
  \caption{Validation of \emph{SeamGain}. (a) Predicted versus measured reduction of global routing wirelength,
  for all configurations of the held-out design. (b) \red{Median reduction versus $\Delta$ at $U=91\%$;} \yl{each point is the median over all swept configurations with that $\Delta$ (lines: predicted; markers: measured)}.
  (c) Best predicted reduction versus the measured reduction of the \emph{SeamGain}-selected configuration, for each design and utilization;
  \yl{RSC is applied if the best predicted reduction is at least $\tau$ (dotted line)}.}
  \label{fig:model}
  \vspace{-1em}
\end{figure*}

\subsection{Source of the Wirelength Reduction}
\label{subsec:source}

\begin{table}[!htb]
\caption{Stage breakdown ($\Delta=2\%$, $U=90\%$--$92\%$; medians over configurations).
$\Delta L_{\rm gp}$: change of global placement HPWL. Red.: reduction of global routing wirelength.}
\label{tab:stages}
\centering
\scriptsize
\setlength{\tabcolsep}{3pt}
\begin{tabular}{|l|r|c|c|c|c|c|c|r|}
\hline
\multirow{2}{*}{\textbf{Design}} & \multirow{2}{*}{$\Delta L_{\rm gp}$} & \multicolumn{2}{c|}{$\lambda_{\rm dp}$} & \multicolumn{2}{c|}{$\lambda_{\rm cts}$} & \multicolumn{2}{c|}{$\rho$} & \multirow{2}{*}{\textbf{Red.}} \\ \cline{3-8}
 & & Base & RSC & Base & RSC & Base & RSC & \\ \hline
\textit{tpu}     & $+0.8\%$ & 1.470 & \textbf{1.392} & 1.190 & \textbf{1.074} & 1.121 & 1.133 & 14.5\% \\ \hline
\textit{ethmac}  & $+1.2\%$ & 1.157 & \textbf{1.140} & 1.455 & \textbf{1.201} & 1.199 & \textbf{1.165} & 16.3\% \\ \hline
\textit{jpeg}    & $+0.8\%$ & 1.335 & \textbf{1.286} & 1.113 & \textbf{1.045} & 1.133 & 1.141 & 7.8\% \\ \hline
\textit{cva6}    & $-1.4\%$ & 1.783 & \textbf{1.392} & 1.282 & \textbf{1.220} & 1.085 & 1.148 & 21.9\% \\ \hline
\textit{swerv}   & $+0.3\%$ & 1.128 & 1.255 & 1.032 & 1.099 & 1.176 & 1.223 & $-19.6\%$ \\ \hline
\textit{mempool} & $-1.1\%$ & 1.182 & \textbf{1.157} & 1.035 & \textbf{1.021} & 1.172 & 1.179 & 2.5\% \\ \hline
\end{tabular}
\end{table}

We validate the analysis of Section~\ref{subsec:mechanism}
using the HPWL reported before and after each
legalization step.
Table~\ref{tab:stages} shows the factors of
Equation~\eqref{eq:stages} for the \yl{uniform flow}
and for RSC with $\Delta=2\%$, \green{the smallest value of $\Delta$ 
that is swept for all six designs.}
We make three observations.
(i) RSC changes the global placement HPWL by only
$-1.4\%$ to $+1.2\%$, as predicted by
Equation~\eqref{eq:cut}.
(ii) In the uniform flow, legalization and CTS inflate
the HPWL by $1.16\times$ to $2.29\times$.
RSC reduces both $\lambda_{\rm dp}$ and
$\lambda_{\rm cts}$ for five of the six designs.
For example, the mean cell displacement of \textit{tpu}
decreases from 0.50$\mu m$ to 0.43$\mu m$ during
legalization, and from 0.28$\mu m$ to 0.12$\mu m$ during
the legalization after CTS.
(iii) The routing factor $\rho$ is not reduced
for five of the six designs.
Consistently, in Table~\ref{tab:main_results} the average
post-route rWL reduction equals that of the
HPWL (15.4\%).
These observations confirm that RSC improves routed
wirelength by reducing legalization- \blue{and CTS-induced} 
wirelength degradation, rather than by reducing routing detours.

The benefit depends on the design.
For \textit{cva6}, which has the largest inflation in the
uniform flow, the reduction is 21.9\%.
For \textit{swerv} and \textit{mempool}, the uniform flow has little
inflation to recover
($\lambda_{\rm dp}\lambda_{\rm cts}\leq1.28$);
the reduction for \textit{mempool} is only 2.5\%, and for \textit{swerv}
the denser blocks increase both inflation factors and
degrade wirelength by 19.6\%.
Hence, RSC should not be applied to every design, which
motivates the design-aware selection of
Section~\ref{subsec:selection}.
Moreover, vertical seams provide most of the benefit.
With $\Delta=2\%$, vertical-only seams \green{($h=0$)} achieve
median reductions of 13.0\%, 16.7\% and 6.0\% for \textit{tpu},
\textit{ethmac} and \textit{jpeg}, and horizontal-only seams \green{($w=0$)} achieve
6.1\%, 8.8\% and $-0.9\%$, compared with 14.5\%, 16.3\% and
7.8\% for both directions.
This is consistent with row-based legalization: a
vertical seam provides free sites in every row.

\subsection{Accuracy of SeamGain}
\label{subsec:accuracy}

\begin{table}[!htb]
\caption{Accuracy of the predicted reduction of global routing wirelength with whole designs held out
(MAE in percentage points; $r$: Pearson correlation). }
\label{tab:accuracy}
\centering
\scriptsize
\setlength{\tabcolsep}{3.6pt}
\begin{tabular}{|l|r|c|c|c|c|c|}
\hline
\multirow{2}{*}{\textbf{Held-out}} & \multirow{2}{*}{\textbf{\#Configs}} & \multicolumn{3}{c|}{\textbf{SeamGain}} & \multicolumn{2}{c|}{\textbf{Ridge}} \\ \cline{3-7}
 & & MAE & MAE (feasible) & $r$ & MAE & $r$ \\ \hline
\textit{tpu}     & 502   & \textbf{5.1} & 2.2 & 0.88 & 6.6  & 0.82 \\ \hline
\textit{ethmac}  & 380   & \textbf{5.2} & 5.4 & 0.42 & 8.8  & 0.42 \\ \hline
\textit{jpeg}    & 490   & \textbf{3.4} & 1.1 & 0.95 & 10.8 & 0.94 \\ \hline
All     & 1,372 & \textbf{4.5} & 3.0 & 0.80 & 8.7  & 0.45 \\ \hline
\end{tabular}
\end{table}

We evaluate \emph{SeamGain} with whole designs held out.
For each of \textit{tpu}, \textit{ethmac} and \textit{jpeg}, we fit the eight
\yl{parameters} on the other two designs, and predict the
reduction of global routing wirelength for all
completed configurations of the held-out design.
The held-out design contributes only its
reference run.
We compare against a ridge regression model on ten
geometric features of the configuration and the floorplan
(e.g., $\Delta/(1-U)$, $w/P_x$, $h/P_y$ and $f$),
trained and evaluated on the same folds.
Global routing wirelength is a reliable proxy for
post-route wirelength: over the 65 detail-routed
configurations, the two reductions have a correlation
of 0.98.

\noindent\textbf{Prediction accuracy. }
Table~\ref{tab:accuracy} and Figure~\ref{fig:model}(a)
show the results.
\emph{SeamGain} achieves a mean absolute error (MAE) of
4.5 percentage points over 1,372 held-out configurations,
compared with 8.7 for the ridge model, and 3.0 over the
packing-feasible configurations ($\delta \leq 0.945$).
For reference, the reduction of a configuration varies
by 1.5 percentage points on average across placement seeds.
The ridge model has no \green{input from the reference 
run and} 
therefore cannot predict the level of the reduction of
a new design: its error for \textit{jpeg} is 10.8.
The fitted \yl{parameters} are stable across the three folds:
$r_d$ is 0.11--0.14, $r_c$ is 0.54--0.60 and $\delta_0$
is 0.948--0.950.
Fitted on all three designs, the \yl{parameters} are
$(r_d, g_d, k_d)=(0.13, 2.7, 21)$,
$(r_c, g_c, k_c)=(0.57, 0.78, 16)$,
$\delta_0=0.950$ and $\kappa=1.013$;
i.e., at the reference pitch, RSC recovers 13\% of the
excess inflation of legalization and 57\% of that of CTS.
For the 65 detail-routed configurations, the
predicted reduction has an MAE of 3.5 against the measured
post-route reduction.

\begin{figure}[t]
  \centering
  \includegraphics[width=0.98\columnwidth]{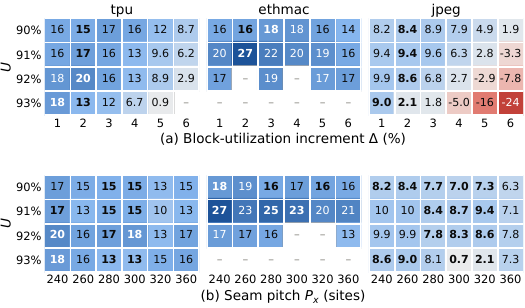}
  \caption{Best wirelength reduction (\%) for each utilization versus (a) $\Delta$ and (b) the seam pitch,
  maximized over the other parameters. Bold: post-route wirelength; regular: global routing wirelength
  (no \red{DRC-}clean detail-routed configuration); ``--'': no completed run or baseline.}
  \label{fig:heatmaps}
  \vspace{-1em}
\end{figure}

\noindent\textbf{Response to the configuration. }
Figure~\ref{fig:model}(b) shows that \emph{SeamGain} reproduces
the response to $\Delta$: the reduction is flat for small
$\Delta$ and \yl{drops sharply} once the packing density exceeds
$\delta_0$.
Over all six designs, none of the 639 configurations with
$\delta\leq0.945$ fails, while 6\% to 22\% fail for
$\delta$ between 0.945 and 0.98; all failures occur in
legalization or are rejected as infeasible packing.
The packing limit thus bounds $\Delta$ by about
$4\%$, $3\%$, $2\%$ and $1\%$ at
$U=90\%$, $91\%$, $92\%$ and $93\%$, which matches the
diagonal boundary of the reductions in
Figure~\ref{fig:heatmaps}(a).
Regarding the pitch, Figure~\ref{fig:heatmaps}(b) shows
that the best reduction is obtained at one of the two
smallest pitches in all 11 cases, as predicted by
the pitch term.
For \textit{ethmac} ($a_{\rm cts}=3.94\%$), the reduction keeps
increasing up to $\Delta=2\%$--$3\%$, as predicted by the
headroom term; however, \emph{SeamGain} underestimates it for
small $\Delta$, since the other two designs do not
exercise this term ($a_{\rm cts}<1\%$).

\noindent\textbf{Unseen designs. }
We further apply \emph{SeamGain}, with the \yl{parameters} fitted on
\textit{tpu}, \textit{ethmac} and \textit{jpeg}, 
to \textit{cva6}, \textit{swerv} and \textit{mempool}, which are
not used in any fit.
For \textit{mempool}, the predicted and measured median reductions
of the packing-feasible configurations are 2.6\% and 2.2\%.
For \textit{cva6}, \emph{SeamGain} is conservative (15.5\% versus 23.0\%).
For \textit{swerv}, it predicts a small reduction (1.9\%),
whereas RSC degrades wirelength; this is the only design
whose sign is not predicted, and the \yl{threshold $\tau$} of
Algorithm~\ref{alg:select} rejects it.
In contrast, the ridge model predicts reductions of
15.1\%--16.5\% for all three designs.
The Rent exponent of the placed designs (0.47--0.58) does
not explain the remaining design dependence: \textit{ethmac} and
\textit{swerv} have the same exponent (0.54) and opposite outcomes.

\subsection{Design-Aware Configuration Selection}
\label{subsec:selection_eval}

\begin{table}[!htb]
\caption{Reduction of global routing wirelength
(averages over utilizations; number of RSC runs per case in parentheses).}
\label{tab:selection}
\centering
\scriptsize
\setlength{\tabcolsep}{2.3pt}
\begin{tabular}{|l|c|c|c|c|c|c|}
\hline
\multirow{2}{*}{\textbf{Design}} & \textbf{SeamGain} & \textbf{Ridge} & \multicolumn{2}{c|}{\textbf{Random}} & \textbf{Median} & \textbf{Best of} \\ \cline{4-5}
 & (1) & (1) & best of 5 & (1) & \textbf{config.} & \textbf{sweep} (144) \\ \hline
\textit{tpu}     & 16.2\% & 16.5\% & 14.9\% & 6.9\%  & 9.6\%  & 18.0\% \\ \hline
\textit{ethmac}  & 16.2\% & 16.2\% & 17.8\% & 14.2\% & 14.5\% & 20.7\% \\ \hline
\textit{jpeg}    & 8.7\%  & 8.3\%  & 7.5\%  & $-1.4\%$ & 2.5\%  & 9.8\% \\ \hline
Average (11 cases) & 13.5\% & 13.5\% & 13.0\% & 5.9\% & 8.3\% & 15.7\% \\ \hline
\end{tabular}
\end{table}

We evaluate Algorithm~\ref{alg:select} with $\Theta$ being
the swept configurations, $\epsilon=0.25$ percentage
points, $\tau=8\%$, and parameters fitted without the design
under test, as in Section~\ref{subsec:accuracy}.
The packing limit $\delta_{\max}$ is the largest packing
density below which no configuration of the fitting
designs fails (0.948).

\noindent\textbf{Selected configurations. }
Table~\ref{tab:selection} compares the \emph{SeamGain}-selected
configuration, which requires a single RSC run, with
reference selections.
Over the 11 cases of \textit{tpu}, \textit{ethmac} and \textit{jpeg} with a completed
baseline, the selected configuration achieves an average
reduction of 13.5\%, i.e., 86\% of the reduction of the
best swept configurations\footnote{\green{With $k$ RSC runs, i.e.,
the $k$ configurations of $\Theta_t$ that
Algorithm~\ref{alg:select} ranks highest, of which the best
measured one is kept, this share increases from 86\% for $k=1$ 
to 89\% for $k=2$ and 93\% for $k=3$.}}, and its predicted reduction
differs from the measured one by 2.1 percentage
points on average.
\blue{Note that the best swept configuration is identified by
its measured wirelength with a single placement seed. When rerun with four other placement
seeds, the best swept configurations achieve a median reduction
of 14.3\% on average; the \emph{SeamGain}-selected configuration,
which is chosen without any RSC run, achieves 94\% of this
reduction.}
It outperforms the best of five random configurations
(13.0\%).
The selection is not sensitive to the tie-breaking rule:
all configurations in $\Theta_t$ complete the flow, and
the worst of them achieves 12.4\% on average.
The ridge model selects equally good configurations
(13.5\%), but cannot decide whether RSC should be applied
(see below).

\noindent\textbf{Applicability. }
Figure~\ref{fig:model}(c) shows the best predicted
reduction and the measured reduction of the selected
configuration for the 20 cases of the six designs that
have a packing-feasible candidate.
The best predicted reduction is at least 9.6\% for \textit{tpu},
\textit{ethmac}, \textit{jpeg} and \textit{cva6}, 
and at most 6.0\% for \textit{swerv} and
\textit{mempool}.
With $\tau=8\%$, Algorithm~\ref{alg:select} applies RSC to
14 cases, all of which are improved, by 7.3\% to 25.5\%;
all 26 available runs of their selected configurations
with different placement seeds have smaller wirelength than
the best of the five baseline seeds.
It keeps the \yl{uniform flow} for the six cases of \textit{swerv}
and \textit{mempool}, for which the selected configuration
achieves reductions of only $-12.0\%$ to $+4.4\%$.
Note that $\tau$ is calibrated on these six designs;
any value between 6.0\% and 9.6\% gives the same decisions.

\section{Conclusion and Future Directions}
\label{sec:conclusion}
We have presented \emph{Reverse Seam-Carving} (RSC),
a \yl{pre-placement} methodology that redistributes
whitespace into core-spanning low-density seams at
fixed die area and utilization.
Our results indicate that RSC reduces routed 
wirelength by mitigating legalization- 
and CTS-induced wirelength degradation.
Experimental results in the OpenROAD flow demonstrate
(maximum, average) post-route wirelength reductions of
(27.6\%, 15.4\%).
From a single reference run of the uniform flow,
\emph{SeamGain} decides whether RSC should be applied,
and selects a configuration that obtains 86\% of the
best-of-sweep reduction.
Ongoing extensions to RSC include:
(i) developing adaptive RSC, a reactive variant that updates 
the seams using placement and routing feedback; 
(ii) exploring the effect of seam-grid offsets and nonuniform seam pitches; \blue{and (iii) evaluating RSC across additional PDKs, commercial P\&R tools, and placement and routing flows with varying optimization capabilities.}

\smallskip

\noindent\textbf{AI disclosure.}
Anthropic Claude 5 and OpenAI GPT 6 models were used
to assist with code and manuscript editing.



\end{document}